\documentclass[conference]{IEEEtran}
\IEEEoverridecommandlockouts
\usepackage{cite}
\usepackage{amsmath,amssymb,amsfonts}
\usepackage{algorithmic}
\usepackage{graphicx}
\usepackage{textcomp}
\usepackage{xcolor}
\usepackage{comment}
\def\BibTeX{{\rm B\kern-.05em{\sc i\kern-.025em b}\kern-.08em
    T\kern-.1667em\lower.7ex\hbox{E}\kern-.125emX}}
\begin{document}

\title{Procedurally generating rules to adapt difficulty for narrative puzzle games
}

\author{\IEEEauthorblockN{1\textsuperscript{st} Thomas Volden}
\IEEEauthorblockA{\textit{Center for Digital Play} \\
\textit{IT University of Copenhagen}\\
Copenhagen, Denmark \\
thvo@itu.dk}
\and
\IEEEauthorblockN{2\textsuperscript{nd} Djordje Grbic}
\IEEEauthorblockA{\textit{Center for Digital Play} \\
\textit{IT University of Copenhagen}\\
Copenhagen, Denmark \\
djgr@itu.dk}
\and
\IEEEauthorblockN{3\textsuperscript{rd} Paolo Burelli}
\IEEEauthorblockA{\textit{Center for Digital Play} \\
\textit{IT University of Copenhagen}\\
Copenhagen, Denmark \\
pabu@itu.dk}
}


\maketitle

\begin{abstract}
This paper focuses on procedurally generating rules and communicating them to players to adjust the difficulty. This is part of a larger project to collect and adapt games in educational games for young children using a digital puzzle game designed for kindergarten. A genetic algorithm is used together with a difficulty measure to find a target number of solution sets and a large language model is used to communicate the rules in a narrative context. During testing the approach was able to find rules that approximate any given target difficulty within two dozen generations on average. The approach was combined with a large language model to create a narrative puzzle game where players have to host a dinner for animals that can't get along. Future experiments will try to improve evaluation, specialize the language model on children's literature, and collect multi-modal data from players to guide adaptation.
\end{abstract}


\section{Introduction}
Research on serious games has shown that when done correctly an educational game can increase a child's motivation to learn and by extension optimize the learning outcome \cite{iten_learning_2016}.

The concept of adaptive learning can be dated back to as early as 1970\cite{carbonell_ai_1970}. The core principle is to track the performance of the learner and to adapt the learning content to improve this performance.

Due to their ability to customize content for the player, digital games have the potential to maximize learning. This potential has been explored in different fields, such as medical training and schools with mixed results \cite{zhonggen_meta-analysis_2019}\cite{villesseche_enhancing_2019}. 
Although the use of serious games to improve cognitive capacities has yielded promising results\cite{plass_effect_2019}, the potential of adapting to the player may not have been fully realized\cite{vanbecelaere_effectiveness_2020}.

One promising direction in the field, multi-modal learning analytics, focuses on fusing multiple sources of data to build more comprehensive learning profiles\cite{mu_multimodal_2020}. This approach has shown the potential to improve the modelling of the learner especially for children. There are however limited studies in the area, due to technical challenges with multi-model data collection for children\cite{crescenzilanna_multimodal_2020}.

Several studies have investigated the interplay between multi-modal player profiling and procedural content generation and adaptation in games\cite{yannakakis_experience-driven_2015}.

A few recent works have explored the application of procedural content generation methods for learning games, but the studies have been confined within on-screen fully digital experiences\cite{hooshyar_data-driven_2018} or on-screen simulation of physical experiences\cite{symeon_retalis_creating_2008}.

This project approaches procedural content generation for physical tile-based games in collaboration with YOLI\footnote{https://www.playyoli.com/} which specializes in tangible digital learning games for children in kindergarten. Specifically, this project case studies the YOLI Board, a physical game device developed by YOLI, which consists of a board and a set of tiles with pictures.
The board contains five spaces for the player to place tiles which the board then registers and interacts with by shaking or throwing off a tile. Games are packaged as a set of thirty tiles and come with various rules and themes to train and support the development of young children.

The digital nature of the YOLI Board allows it to collect anonymous player behaviour information and has the potential to adapt to such information. This makes it a promising candidate as a test bed for dynamic difficulty adjustment. Even with the constraints of a fixed set of tiles, it is possible to adapt the difficulty by complicating or simplifying rules.

The goal of this project is to answer this research question. 
How can player behaviour be gathered and used to procedurally generate content and adapt the learning experience to improve the social and cognitive skills of children?
This paper will focus on the procedural content generation part of the research question and try to answer a more specific question in order to contribute to the final answer.
How can a game with a fixed set of tiles designed for kindergarteners dynamically adapt difficulty by changing and communicating rules in such a way that the children are challenged but not overwhelmed?

The term rules is used despite the fact that the examples are limited to constraints that may be considered level design for a puzzle game, such as adding numbers to a Sudoku problem to limit the solution space.
This is because the platform and procedural generator can theoretically support rules that require player adaptation, such as requesting the player to prevent the board from flinging off a specific tile.

It is essential to recognize that the rules themselves, as well as how they are communicated, add to the difficulty level of a game.

Recent studies have shown promising results by utilizing large language models to produce contextual rationalizations with the use of special prompt techniques\cite{park_generative_2023}\cite{du_guiding_2023}.
This project will use large language models to experiment with different ways to communicate rules and how to use narrative context to rationalize rules that may otherwise be hard to comprehend.

\section{YOLI Board and simulator}

\begin{figure}
    \centering
    \includegraphics[width=80mm]{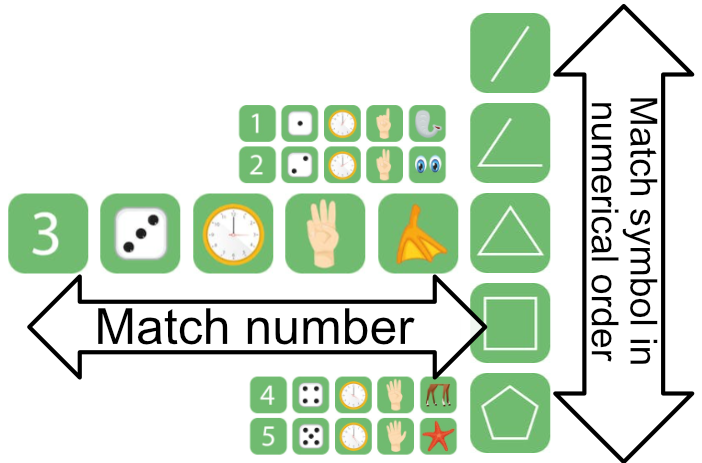}
    \caption{Example of a tile game with thirty tiles, where a player has to match tiles that represent the same number or the same symbol in numerical order.}
    \label{fig:tilegame}
\end{figure}

The YOLI company has specialized in child-friendly educational games and has delivered numerous YOLI Boards to kindergartens in the north of Europe.
Under a child-friendly exterior, the YOLI Board is equipped with wireless chip readers to read and match digitally imprinted information on the tiles. The board is able to shake or entirely throw off individual tiles using magnets, which produces tangible feedback for the players.

The goal of most games is to place five tiles on the board that complies with the rules, which are mainly to match certain attributes of the tiles. An example of this is the "Fun with numbers" game, shown in figure \ref{fig:tilegame}, which features thirty tiles representing a number. The rule of the game is to match tiles that represent the same number or match tiles of the same symbol in numerical order.

The board can connect with a mobile phone where a dedicated app can assist with audio and other more advanced game features. 
The mobile app is also used for data collection and transfer from the YOLI Board.

A software simulator of the physical board had to be developed to test and evaluate games and rules efficiently. The simulator was developed in Python and can be operated manually or via the gym environment for training and testing agents~\cite{brockman_openai_2016}.

The simulator was built around the architectural decision to separate the board and games, which means that the board simulation can be initialized with games that inherit from a tile game-specific abstract class. The board implementation maintains state information above what tiles are placed where and handles user input and game evaluation output. Tile games provide a list of tiles and a method to evaluate a solution set of tiles and return a list to accept, reject or ignore individual tiles.

\section{Procedural rule generator}
The simulator provides tools to assist in the creation of custom tile-based games and accommodate game adaptations with procedural rule generators and rule mutations for genetic algorithms.

The procedural rule generator generates rules based on rule concepts and parameters. A rule concept should limit the number of solutions with precision and be formulated in a way that can reflect a conflict in a narrative context. To accomplish this the following rule concepts were defined, with some designed to exclude a large number of tiles while others exclude a small but specific set of tiles:
\begin{itemize}
    \item Exclude any where \emph{property} is equal to \emph{value}.
    \item Exclusively for any where \emph{property} is equal to \emph{value}.
    \item Only the ones where the value of \emph{property} match.
    \item Those with \emph{property} set to \emph{value} cannot be adjacent to those with \emph{property2} set to \emph{value2}.
    \item There can only be \emph{number} with \emph{property} set to \emph{value}.
\end{itemize}
   
The procedural rule generator picks a rule concept from the list at random and samples the parameters (\emph{property} and \emph{value}) from a list of tiles, like the latest set of tiles that solved the puzzle.
For example, a game featuring vehicle tiles could sample the property \emph{type} and a \emph{value} among \emph{car}, \emph{bus}, or \emph{motorcycle}. Combined with the rule concept this can form the rule "Vehicles of type bus can not be next to vehicles with type motorcycle".

Multiple rules can be combined to add more precision. This is done using a special rule concept called a composite rule, to which child rules can be added or removed. The composite rule evaluates a solution on each of its child rules and merges the result using pair-wise comparison. A tile is rejected if at least one of the child rules rejects the tile.

The difficulty is determined by the number of allowed solutions, consisting of sets of five tiles that are not rejected by the rules. With a maximum of 142,506 solutions for a 30-tile game where the order is irrelevant, this measure indicates an increasing challenge level, from trivial to impossible, as the number of solutions decreases.
The evaluation of difficulty and procedural rule generator have been combined in a genetic algorithm to search for a target difficulty. 

Rule concepts are mutated by changing one or more random parameters for the rule concept. The composite rule concept is also able to add and remove a random rule as part of its mutation.

\section{Framing rules in a narrative context}
Procedurally generated rules, such as the vehicle example given earlier, naturally result in the question ``why?''.
To minimize this, a large language model is used to communicate rules, which at the same time allows for more creative delivery than the rule generator can provide.
The current approach assembles a prompt that contains information about the game and a request to generate a story that can contextually explain the rules without mentioning them explicitly.


OpenAIs ChatGPT 3.5-turbo API\cite{openai_no_nodate} is currently used to prompt for a story, which, when specifically instructed to, can produce children's stories. The model has occasionally used words that are not exactly child friendly, which may be avoided with further fine-tuning of the prompts.

\section{Current results}
To test the procedural content generator combined with the genetic algorithm, repeated tests were conducted five hundred times with a fixed population size of one hundred rules and a mutation rate of fifty percent. For each test, a unique target between zero and all possible solution sets was chosen, and a maximum of fifty generations. The algorithm was able to approximate the target set of solutions with 99.9 percent on average with a standard deviation of 0.17 percent within an average of 22.3 generations and a standard deviation of 16.6 generations.

\begin{figure}
    \centering
    \includegraphics{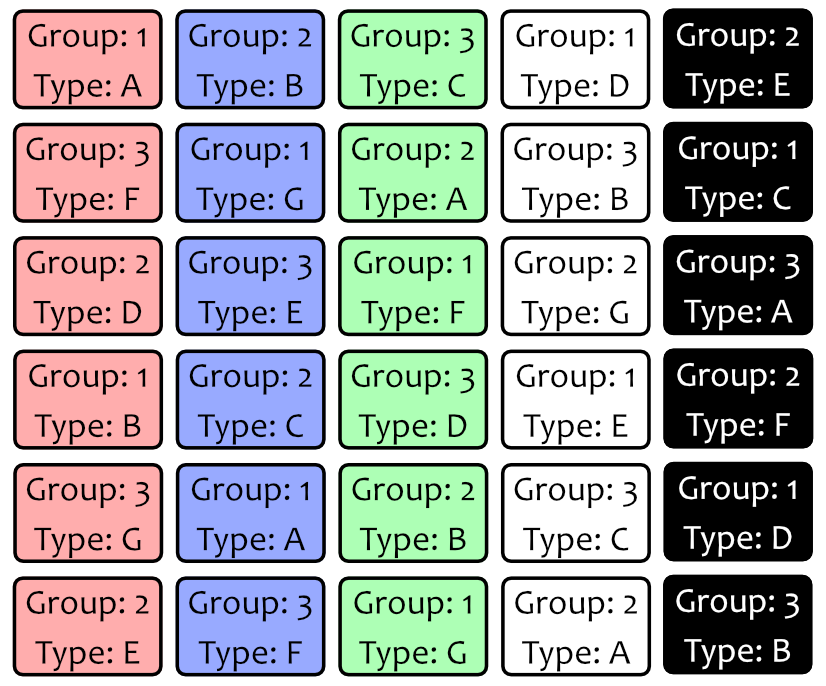}
    \caption{Thirty generic tiles generated, with the properties group, type, and color, to use as a sample for the random rule generator.}
    \label{fig:generic-tiles}
\end{figure}

Given the constraining nature of the rules, thirty generic tiles were generated with three properties, shown in figure \ref{fig:generic-tiles}, each with a different value count (three, five, and seven). The three properties and values were \emph{group}, which had three values (1,2,3), \emph{color} which had five values, and \emph{type} which had seven values (A, B,..., G). These were sampled during initialization by the random generator and again during mutation, which allowed for more dynamic outcomes for the procedural rule generator. 

The difficulty evaluation is the main bottleneck for the performance of the genetic algorithm as it uses brute force for all possible combinations on the entire population.

Here are three samples with different targeted solution sets to give a sense of the rules that the procedural generator outputs. For an almost trivial problem (141,966 solutions) the rules limit tiles which there are only a limited number of: "There can only be 3 with type set to A and There can only be 3 with color set to red". The first rule only affects solutions where more than 3 of the 5 tiles with type A are present and the second does the same for solutions where 3 of the 6 tiles with the color red are present.

For a constrained problem (186 solutions), the rules tend to exclude many tiles by targeting a common property: "Exclusively for anywhere group is equal to 1". Since there are only three group values, this effectively excludes ten tiles entirely.
Between these two extremes (70,764 solutions), the more complex rule concepts tend to be combined: "Those with group set to 3 cannot be adjacent to those with type set to E and Those with color set to green cannot be adjacent to those with color set to white"
These rules partially exclude some tiles when other tiles are present in the solution which combined with common properties such as group and color provide finer-graded control of the solution sets.

\begin{figure}
    \centering
    \includegraphics{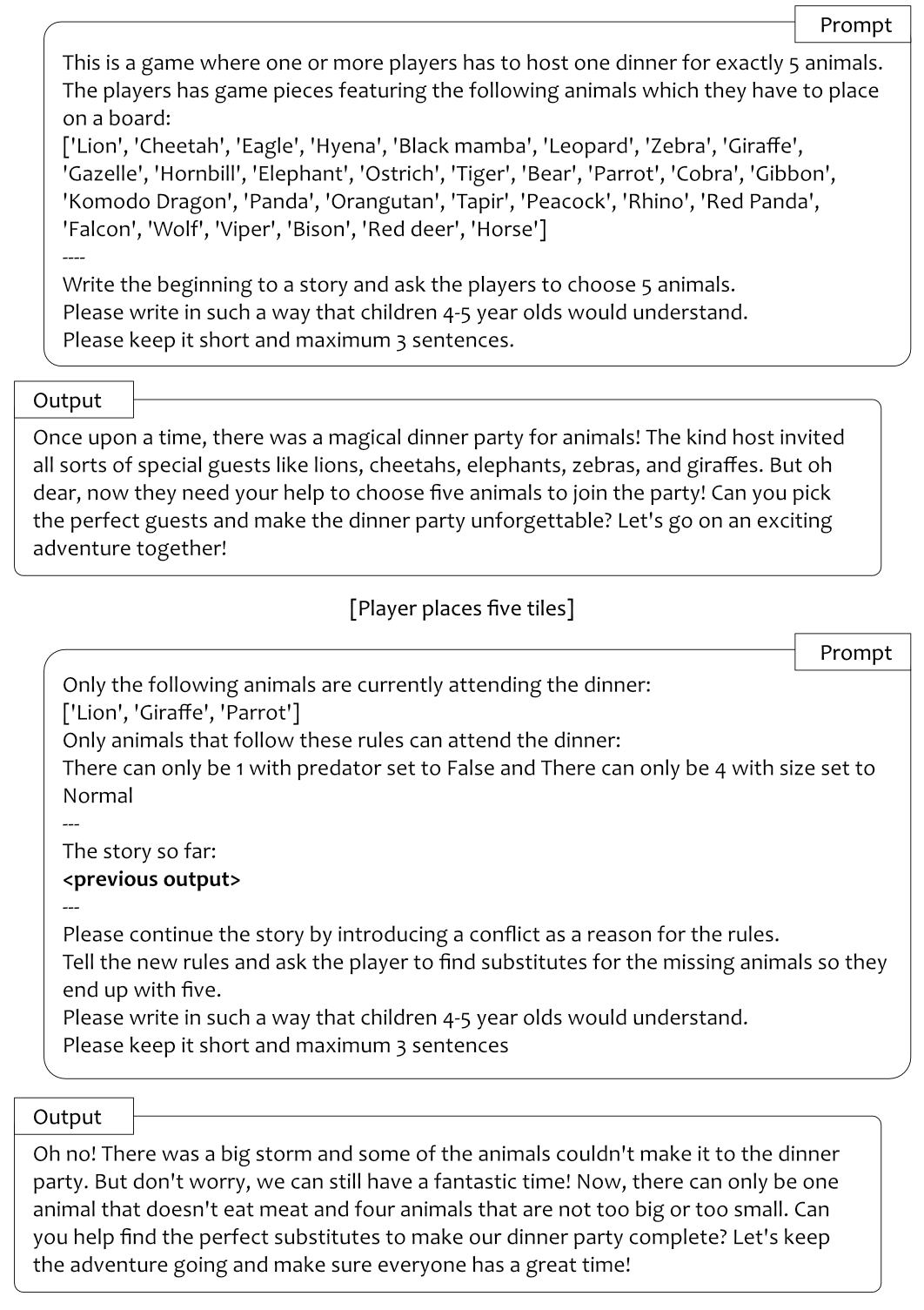}
    \caption{Prompt input to generate a story from an initial game definition and new prompt after the player has placed five tiles. The outputs are from the language model.}
    \label{fig:story-context}
\end{figure}

For the language model to provide a narrative context the tiles had to feature themed entities, which can be incorporated in a story and has a means to conflict. This resulted in a game where the players have to host a dinner for various animals, where individual animals have conflicting constraints that the player has to abide by.
Before the large language model is utilized a prompt is compiled, shown in figure \ref{fig:story-context}, consisting of a game definition where all tiles are mentioned by name and instructions to generate a story and ask the player to choose five animals. A similar prompt is processed every time new rules are procedurally generated, with instructions to continue the story.


\section{Future plans}
In the next iteration of the project, we plan to train a language model specifically on children's literature, provided by the Gutenberg project (gutenberg.org).
This should produce more children-themed stories and hopefully use a language that children are more familiar with.

Large language models have been used to guide reinforcement learning with the use of prompting techniques and evaluation of text output \cite{du_guiding_2023}.
We will try to use large language models to analyze the communicated rules and use prompting techniques to get suggestions for possible solutions.
These suggestions will then be used to guide a solver algorithm and at the same time determine comprehensive difficulty based on the accuracy of the predictions.

The number of unique solutions is a temporary guide for difficulty, but may not be the most fitting evaluator.
Alternatively, entropy has been used as a measure for game depth in strategic games\cite{lantz_depth_2017} and may potentially be used to determine resource requirements. It is currently possible to measure the entropy for a given board state by counting the number of tiles that can fit each board position. This has been used as an evaluator together with the genetic generator to produce rules with a target entropy level between high and low. However, this needs to be studied further in future iterations.

The current approach also lacks actual player feedback on difficulty, which will be an important factor to balance and guide future experiments.
Players will be invited to play test different prototypes, while data is collected to model their behavior in response to different levels of difficulty.
This will be used to evaluate different guiding evaluators for the procedural content generator as well as different language models' ability to communicate game rules.

\bibliographystyle{plain}
\bibliography{references}

\end{document}